\documentclass[twocolumn,prd,nofootinbib]{revtex4}

\newcommand\ForInternalReference[1]{}
\newcommand\SkipForEarlyCirculation[1]{}

\newcommand\SkipPP[1]{}
\usepackage{verbatim}
\usepackage{graphicx}
\usepackage{dcolumn}
\usepackage{bm}
\usepackage{color}
\usepackage{xspace}
\usepackage{url}
\usepackage{amsmath}
\usepackage{float}
\usepackage{multirow}
\usepackage{amssymb}
\usepackage{times}
\newcommand\optional[1]{}

\newcommand\unit[1]{{\rm #1}}

\usepackage{color}
\definecolor{amber}{rgb}{1.0, 0.75, 0.0}
\definecolor{orange}{rgb}{1.0, 0.5, 0.0}
\definecolor{amaranth}{rgb}{0.9, 0.17, 0.31}

\graphicspath{{./figures/}}

\def\ltsima{$\; \buildrel < \over \sim \;$}
\def\simlt{\lower.5ex\hbox{\ltsima}}
\def\gtsima{$\; \buildrel > \over \sim \;$}
\def\simgt{\lower.5ex\hbox{\gtsima}}

\def\RIT{Center for Computational Relativity and Gravitation, Rochester Institute of Technology, Rochester, New York 14623, USA}

\begin{document}
\renewcommand{\arraystretch}{1.5}
\title{Eccentricity matters: Impact of eccentricity on inferred binary black hole populations}

\author{M. Zeeshan}
\email{m.zeeshan5885@gmail.com}
\affiliation{\RIT}

\author{R. O'Shaughnessy}
\email{richardoshaughnessy.rossma@gmail.com}
\affiliation{\RIT}

\begin{abstract}
Gravitational waves (GW), emanating from binary black holes (BBH), encode vital information about their sources 
enabling us to infer critical properties of the BBH population across the universe, including mass, spin, 
and eccentricity distribution. While the masses and spins of binary components are already recognized for their 
insights into formation, eccentricity stands out as a distinct and quantifiable indicator of formation and 
evolution. However, despite its significance, eccentricity is notably absent from most parameter estimation (PE) 
analyses associated with GW signals.
To evaluate the precision with which the eccentricity distribution can be deduced, we generated two synthetic
populations of eccentric binary black holes (EBBH) characterized by non-spinning, non-precessing dynamics, and 
mass ranges between $10 M_\odot$ and $50 M_\odot$. This was achieved using an eccentric power law model, 
encompassing $100$ events with eccentricity distributions set at $\sigma_\epsilon = 0.05$ and 
$\sigma_\epsilon = 0.15$. This synthetic EBBH ensemble was contrasted against a circular binary black holes
(CBBH) collection to discern how parameter inferences would vary without eccentricity. Employing Markov 
Chain Monte Carlo (MCMC) techniques, we constrained model parameters, including the event rate 
($\mathcal{R}$), minimum mass ($m_{min}$), maximum mass ($m_{max}$), and the eccentricity
distribution ($\sigma_\epsilon$). Our analysis demonstrates that eccentric population inference can identify 
the signatures of even modest eccentricity distribution. In addition, our study shows 
that an analysis neglecting eccentricity may draw biased conclusions of population inference for 
the larger values of eccentricity distribution. 
\end{abstract}
\maketitle

\section{Introduction}

Black holes exist throughout the universe in different sizes and types \cite{Frolic_BH_Book_2011}. 
When two black holes become gravitationally bound, they form a binary and start coalescing 
\cite{Peters-1963}. When they coalesce, they produce ripples in the fabric of space-time, known as GW \cite{Frolic_BH_Book_2011}. 
These waves are the primary way to detect those mergers in space, 
starting with LIGO's \cite{LIGO_2015} first detection of a BBH merger (GW150914). Afterward, LVK network 
\cite{LIGO_2015, VIRGO_2012, Virgo-2015,2021PTEP.2021eA101A} continued to regularly detect gravitational waves from 
various compact binaries such as BBH, binary neutron stars (BNS), and binaries composed
of a neutron star and black hole (NSBH) \cite{2019PhRvX...9c1040A,2021PhRvX..11b1053A,2021arXiv210801045T,2023PhRvX..13d1039A}. These waves carry 
information about the progenitor, such as component mass, spin, period, eccentricity, distance, and location.
Indirectly, these properties may provide clues into how binaries formed. These binaries are believed to form from two formation scenarios \cite{Mandel-2022}, 
isolated and dynamic. The isolated binaries are thought to result from the death of binary stars, either 
by a supernova or a common envelope \cite{Mapelli-2020,steinle-2021}.  On the other hand, two black holes
can also make a binary through dynamic encounters in dense star clusters \cite{Antonini-2016,antonini-2019}.
Importantly, each formation scenario imprints on binary parameters such as mass, spin, and eccentricity. 
For instance, isolated binaries tend to have circular orbits in the LVK band due to mass transfer and most
probably the result of gravitational decay over time.  On the contrary \cite{Mapelli-2020}, the dynamically 
formed binaries may have eccentric orbits.

Mass and spin are widely used to understand a binary's properties, evolution, and formation channel. 
However, many of these same formation scenarios also can introduce eccentricity in BBH orbits, such as 
stellar scattering, dynamic interactions in dense environments, or interaction of a third object with a binary.
Furthermore, low energy and slow-forming binaries usually start with modest eccentricities but at very low frequencies; 
over many decades of frequency evolution, gravitational wave radiation circularizes their orbits 
\cite{Peters-1964} before entering the LVK frequency band. However, binaries formed in more violent, 
energetic environments, such as globular clusters, can be formed closer to the LVK frequency band and thus 
retain more of their large natal eccentricities when observationally accessible.   Previous studies have 
demonstrated that identifying orbital eccentricity could differentiate between formation channels 
\cite{Rod-2018,zevin-samsing-2019,samsing-2018, Rodriguez-2018,Antonini-2014}. 


Although eccentricity is a signature that can be measured with GW and is unique to extreme events such as 
high mass ratio mergers, all the confirmed detections so far may be consistent with a nearly circular orbit
in the LVK frequency band \cite{Isobel-2022,hector-jake-2022}. However, some investigations suggest that 
BBHs like GW190521 may exhibit some indications of eccentricity 
\cite{Gamba_2022_GW190521_dynamical,yumeng-2023, 2023NatAs...7...11G, 2020arXiv200905461G,2020ApJ...903L...5R}.
On the other hand, we do have studies \cite{2022arXiv220801766I,2023PhRvD.108l4063R}which do not find the evidence of eccentricity in GW190521. So, it's 
still an exciting event with question whether eccentricity is present in the GW190521 or not.
Unfortunately, these inferences rely on incompletely-surveyed waveforms produced either by direct simulation 
or (more customarily) by phenomenological approximations tuned to those simulations.  Currently, available
phenomenological models allowing for eccentricity only cover part of the possible parameters: non-precessing
binaries, for example. 
While researchers are actively producing the eccentric waveform models analytically \cite{Huerta-2014} and 
numerically \cite{gold-2016,hinder-2010, Healy-2022, Alessandro-2022, Campanelli-2009}, at present, parameter
inference capabilities are limited. 

Despite severe limitations on single-event parameter inference, a few proof-of-concept studies have 
investigated how to identify the presence of an eccentric subpopulation from observations of many massive
binary black holes \cite{lower-2018-ecc-pop, Fang-2019,wu-2020}.
The techniques used in these investigations borrow from extensive studies on reconstructing the population 
properties of quasi circular binaries over the whole mass spectrum
\cite{Dan_2019, Michael-2015, samsing-hamers-2019, 
belczynski-2016, Colm-2017, Akinobu-2017, farr-2017-nature, Richard-2017-natal-kicks, 
Dan-Richard-2018, Abbot-2019-pop, 2021ApJ...921L..43Z}.
Because eccentricity is so poorly constrained by short GW observations of massive BH binaries, these studies
find eccentricity can only be resolved with great difficulty, even with many observations.

By contrast, for low-mass binaries, previous studies have demonstrated that eccentricities can be particularly
well-constrained for low-mass objects \cite{kate2024}, owing to their long modulated inspiral \cite{favata-scaling-2022},
even it improves the accuracy in parameter inference \cite{2015PhRvD..92d4034S}. We use this parameter inference investigation as our prototype for the synthetic inferences about synthetic
GW sources used in our proof-of-concept study.

In this study, we focused on the non-spinning, non-precessing, lower mass $(10 M_\odot - 50 M_\odot)$.
In addition, we will use mass ratio $q = m_1 / m_2 $ with condition 
$m_1>m_2$ and total mass $M = m_1 + m_2 = 100 M_\odot$. We also infer how well we can recover the event rate, mass, 
and eccentricity distribution using the $100$ eccentric events. To extend our analysis, we compared 
constrained parameters using the EBBH and CBBH, which show a considerable difference in recovery of true parameter
for higher eccentric $(\sigma_\epsilon = 0.15)$ population and similar recovery for
lower eccentric population $(\sigma_\epsilon = 0.15)$.

In section \ref{sec:methods}, we described the Bayesian statistical methods used to make the population 
inference. Briefly, we also described the volume-time estimate to accommodate the LVK sensitivity. We 
modified the previously constructed \cite{fishbach-2017,2018talbot_bbh_model} 
power law model to include eccentricity using a truncated normal distribution. We used 
this model to generate a synthetic population and then made 
the inference using MCMC method. Section \ref{sec:syn_pop} describes how we have created a synthetic population using
the eccentric power law model and then added an error in each event to make the population closer to real events 
detectable by LVK. Secondly, we explained the scaling to remove the eccentricity from the synthetic events 
to compare the different constraints using EBBH and CBBH populations. Section \ref{sec:pop_inference} discusses 
our results and explains how well we have constrained and recovered the parameters. Also, their accuracy and 
recovery comparison among EBBH and CBBH. Finally, we summarize our findings in Sec. \ref{sec:conclude}.

\section{Methods}
\label{sec:methods}

A coalescing BBH can be completely described by three intrinsic and seven extrinsic parameters. The intrinsic
parameters, such as the mass of the binary component $(m_i)$, spin $(\chi_i)$, and eccentricity $(\epsilon)$  
determine the orbital evolution of the binary. The extrinsic parameters determine the merger's space-time 
coordinates and orientation. We used the following method to infer the population properties of binary black 
holes.

\subsection{Hierarchical Bayesian Modeling}

We use hierarchical Bayesian modeling (HBM) to constrain a population model with gravitational wave data. 
In HBM, we have $N$ number of discrete detections. Those detections provide merger data denoted as 
$d_1,d_2,d_3,...,d_N$ where each $d_i$ shows a BBH merger. Each individual stretch of data $d_i$  can
be used to infer the properties of the BBH associated with that data segment. These properties, often 
called parameters, are denoted by $\lambda_1,\lambda_2,\lambda_3,...,\lambda_i$. Each parameter has
its uncertainty, and we express it by the probability of the data given the parameter value. We also 
refer to it as the likelihood function $\mathcal{L}(\lambda)=p(d|\lambda)$ of a source. When 
calculated in full with data and a waveform model, the full likelihood  function
expresses the probability of a specific waveform model with parameters lambda in the data $d$. 
Once we have a likelihood function, we may use a uniform prior or any informative prior to find a posterior 
probability using the Bayes theorem as given in Eq. \ref{eq:Bayes_ind}

\begin{equation}
\label{eq:Bayes_ind}    
p(\lambda|d) \propto p(d|\lambda) p(\lambda).
\end{equation}

This posterior probability will constrain the properties of each binary, such as mass, spin, and eccentricity.
We may infer those parameters using rapid parameter inference on gravitational wave sources via iterative 
fitting (RIFT): an open source code for parameter estimation (PE) of the binary sources \cite{rift_2018}.
The above discussion describes the true exact likelihood.  However, as discussed below in section
III.A, following previous work \cite{2017MNRAS.465.3254M} we will employ a synthetic likelihood model instead of performing an 
end-to-end parameter inference calculation.

\subsection{Bayesian Inference}

Now having the PE of individual sources, we follow the Bayesian framework for population inference.
The likelihood of a population parameter $\Lambda$ is equivalent to the probability of the individual 
sources given the population parameter $\Lambda$ is written as follows.

\begin{equation}
\label{eq:likelihood_pop}    
\mathcal{L}(\Lambda)\equiv p(d_1,d_2,d_3,...,d_N|\Lambda).
\end{equation}

We use the likelihood provided in Eq. \ref{eq:likelihood_pop} in the Bayes theorem defined in Eq. \ref{eq:Bayes}
to find posterior probability.

\begin{equation}
\label{eq:Bayes}    
p(\Lambda|d_1,d_2,...,d_N)= \frac{p(\Lambda)p(d_1,d_2,...,d_N|\Lambda)}{p(d_1,d_2,...,d_N)},
\end{equation}
where $p(\Lambda|d_1,d_2,d_3,...,d_N)$ is posterior, $p(\Lambda)$ is prior, and $p(d_1,d_2,d_3,...,d_N)$ is 
normalization constant or also known as evidence.

To conduct our analysis for mass and eccentricity distribution, we will use the inhomogeneous Poisson process
\cite{2019MNRAS.486.1086M, 2004AIPC..735..195L}
scaled by rate $\mathcal{R} = \frac{dN}{dtdV_c}$ and parameterize by $\Lambda$ to find the likelihood 
$\mathcal{L}(\mathcal{R},\Lambda)\equiv p(D|\mathcal{R},\Lambda)$ of an astrophysical population given the
merger rate and parameter $\Lambda$. 

\begin{equation}
\label{eq: likelihood}
\mathcal{L}(\mathcal{R},\Lambda) \propto e^{-\mu(\mathcal{R},\Lambda)} \prod_{n=1}^N\int d\lambda~\ell_n(\lambda)~
\mathcal{R}~ p(\lambda|\Lambda),
\end{equation}
where $\mu(\mathcal{R},\Lambda)$ is the expected number of detection under the given population parametrization 
$\Lambda$ with the overall rate $\mathcal{R}$. $\ell_n(\lambda)=p(d_n|\lambda)$ is the likelihood of the data 
$d_n$ given binary parameter.
Finally, we will get our posterior as follows:
\begin{equation}
p(\mathcal{R},\Lambda | D)\propto p(\mathcal{R},\Lambda)  \mathcal{L}(\mathcal{R},\Lambda)  
\end{equation}
by choosing a appropriate prior $p(\mathcal{R},\Lambda)$.

These calculations are analytically intractable and must be performed numerically. Specifically, we will use 
Goodman and Weare's affine invariant Markov chain Monte Carlo (MCMC) \cite{mcmc_paper} to find the posterior 
distribution of population parameters. This method draws samples from the targeted distribution for $\Lambda$, 
in our case, its eccentric power law model given in Eq. \ref{eq:plawg}, then compares it with the given data 
(collection of individual events) and stores the best-fit sample. We may iterate this as we need and store 
multiple sample values until they converge. 
Our specific implementation is a Python package called EMCEE \cite{emcee_paper}. We use a low-dimensional 
model; chains converge to a common region after a few hundred steps. Therefore, we performed our analysis 
with $100$ walkers up to $5000$ samples and burned the initial $2500$ steps at the plotting stage. 

\begin{figure*}[!htbp]
    \includegraphics[width=0.45\textwidth]{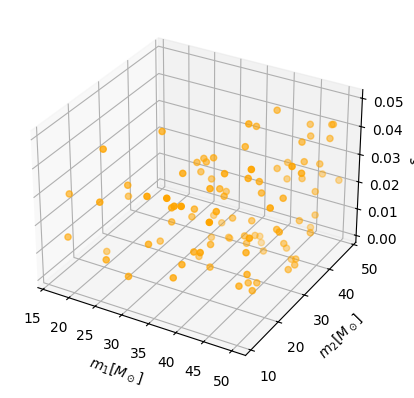}
    \includegraphics[width=0.45\textwidth]{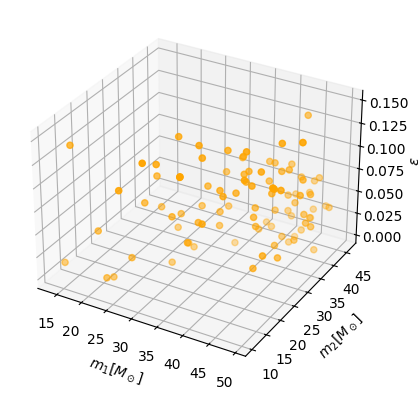}
    \caption{\label{fig:pop3d0.05_0.15} Synthetic Population of EBBH: LHS shows with $\sigma_\epsilon=0.05$, RHS 
    Shows with $\sigma_\epsilon=0.15$}
\end{figure*}

\subsection{Volume Time (VT) Estimation}

To make our study realistic, we include the sensitivity of the LVK instruments. This sensitivity is defined 
by time-volume to which a census of gravitational wave events is sensitive: inferring the product $VT$.  In 
this expression, $V$ is the characteristic volume with units $Gpc^{3}$, which refers to the possible detection 
region in the sky for the LVK \cite{Volume_1993}, and $T$ is the time duration of making observations at this 
sensitivity.  In practice, $VT$ reflects a suitable time-averaged or cumulative sensitivity, as the true 
network and sensitivity vary over time.

Existing LVK instruments' sensitivity depends primarily on the mass and, to a lesser extent, on binary spin and
(if present) modest eccentricity.  Since we neglect spin in this work, we assume the network will have the 
same  VT versus mass as was previously estimated  \cite{Dan_2019} for non-spinning, non-eccentric, and 
non-precessing binaries.  Hence, we briefly explain the calculations; see \cite{Dan_2019} for details. The 
Eq. \ref{eq:volume} calculates the orientation averaged sensitive volume \cite{Abbott_2016,richard2010volume}

\begin{equation}
\label{eq:volume}
V(\lambda) = \int P((<D(z))/D_h(\lambda))\frac{dV_c}{dz}\frac{dz}{1+z},
\end{equation}    
where $D(z)$ is the luminosity distance for redshift $z$, $D_h$ is the horizon 
distance to which source can be seen, and $V_c$ is the comoving volume 
(for details see \cite{Dan_2019}).  
Finally, to compute the average number of detections, we used Eq. \ref{eq:mu}.

\begin{equation}
\label{eq:mu}
  \mu(\mathcal{R},\Lambda) = \int(VT)~\lambda ~ \mathcal{R}~p(\lambda|\Lambda)~d\lambda ,
\end{equation}
where $p(\lambda|\Lambda)$ is the probability density function for a random binary in the universe to have 
parameter $\lambda$. Keep in mind that $\lambda$ is equal to all intrinsic and extrinsic parameters.

\subsection{Eccentric Power Law}

There are various weak and pure phenomenological population models proposed in previous studies 
\cite{2016PRXAbbot_BBH_model,2017FishBach_BBH_model,2018talbot_bbh_model}. We don't have 
enough well-understood theory encompassing all possible formation channels to choose any
specific model. Therefore, our analysis used the pure truncated power law defined in 
\cite{2016PRXAbbot_BBH_model,2017FishBach_BBH_model} and modified it to 
include one-sided Gaussian eccentricity. This model computes the intrinsic probability 
of $m_1$, $m_2$, and $\epsilon$.
For simplicity, to focus on binaries with some degree of pre-merger signal, not necessarily to
encompass all astrophysical population models or even rare extreme detections, we kept 
total mass $M_{max}=m_1+m_2 = 100 M_\odot$. We also assume that non-zero probability density
only exists for $m_{min}\leq m_2 \leq m_1 \leq m_{max}$.
The generalized form of the truncated power-law model with parameters 
$\Lambda \equiv  (\alpha, \mathcal{R}, k_m, m_{min}, m_{max}, \sigma_\epsilon, M_{max})$ and random variable 
$m_1$, $m_2$, and $\epsilon$ has the functional form in Eq. \ref{eq:plawg} within provided mass limit.

\begin{align}
\label{eq:plawg}
p(m_1,m_2,\epsilon) = &C(\alpha,k_m,m_{min},m_{max},M_{max},\epsilon)  
\nonumber \\ & \sqrt{\frac{2}{\pi}} \frac{(m_2/m_1)^{k_m} m_1^{-\alpha} e^{-(\epsilon/\sqrt{2}
\sigma_\epsilon)^2}}{(m_1-m_{min})\sigma_\epsilon},
\end{align}
where $\alpha$ is the power law index, $\mathcal{R}$ is the merger rate, $m_{min}, m_{max}$ are the minimum 
and maximum masses of the binary components in the population, and $\sigma_\epsilon$ is the orbital 
eccentricity distribution. The Eq. \ref{eq:plawg} represent a truncated power law for primary mass $m_1$ with 
index $-\alpha$ and conditional power law distribution $p(m_2|m_1)$ for secondary mass $m_2$ using power law, 
and one-sided Gaussian distribution for orbital eccentricity $\epsilon$. 
For our analysis, we defined a constant of integration equal to $\int_V dm_1 dm_2 d\epsilon p(m_1,m_2,\epsilon) 
= 1$.  Our detectors are sensitive to high-mass BBHs, therefore, we will
use $k_m=0$ throughout the studies. As a result, we have our reduced form of the truncated power law in 
Eq. \ref{eq:plaw}

\begin{align}
\label{eq:plaw}
p(m_1,m_2,\epsilon) = \sqrt{\frac{2}{\pi}} \frac{ m_1^{-\alpha}  e^{-(\epsilon/\sqrt{2}\sigma_\epsilon)^2}}{(m_1-m_{min})\sigma_\epsilon}.
\end{align}

\section{Synthetic Population}
\label{sec:syn_pop}
We have generated two synthetic populations with conservative case $\sigma_\epsilon = 0.05$ and optimistic
case $\sigma_\epsilon = 0.15$ by choosing the power law parameters $\alpha = -1$, $m_{min} = 10$, $m_{max}=50$.
Starting with  $10000$ synthetic sources for each population, we find the probability for each event to be detected
by computing the VT of 
each source. Finally, we weighed based on these VTs and randomly picked $N=100$ sources from each population to 
perform our analysis.  To be concrete, to eliminate the impact of Poisson counting statistics on our event rate
inference and guarantee that both synthetic populations should favor the same event rate, we have specifically 
adopted a known event count for each population (i.e., $N=100$).  As a result, the
implied observing time ($T=248\unit{d}$) guarantees the expected event count will be exactly $\mu=100$ for the true
population parameters for each synthetic population. In our analysis, we assumed the fixed rate density 
per comoving volume, and no evolution.
Our injected populations with $\sigma_\epsilon=0.05$ and
$\sigma_\epsilon=0.15$ are shown in Fig. \ref{fig:pop3d0.05_0.15}.
\begin{figure*}[!htbp]
    \includegraphics[width=0.45\textwidth]{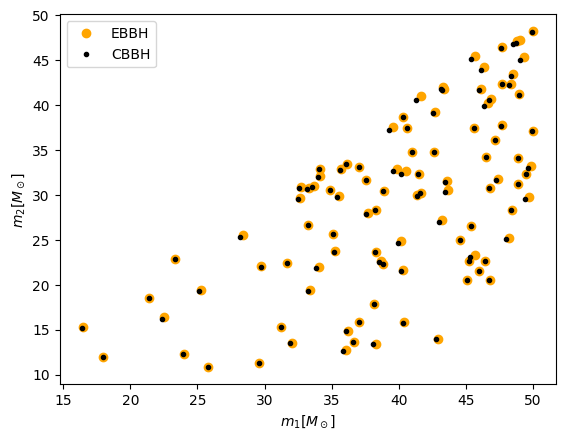}
    \includegraphics[width=0.45\textwidth]{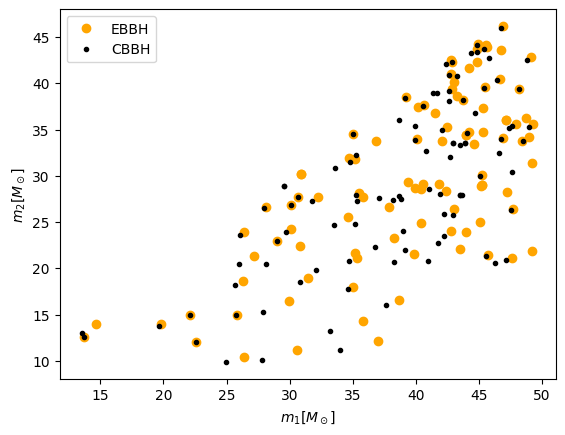}
    \caption{\label{fig:pop2d_0.05_0.15} Mass Shift after Removing Eccentricity: LHS Shows with 
    $\sigma_\epsilon =0.05$, RHS Shows with $\sigma_\epsilon=0.15$}
\end{figure*}

\begin{table*}
    \centering
    \begin{tabular}{c|ccccc}
        \hline \hline
       Quantity & $\log_{10}(\frac{\mathcal{R}}{\rm Gpc^{-3}yr^{-1}})$ & $\alpha$ & $m_{min} [M_\odot] $ & 
       $m_{max} [M_\odot]$ & $\sigma_\epsilon$ \\ \hline
      Synthetic population & 2 & -1 & 10 & 50 & \{0.05, 0.15\} \\ \hline
      Prior range & $[-5,5]$ & $[-5,5]$ & $[1-20]$ & $[30-100]$ & $[0-0.5]$ \\ \hline
      Prior distribution & Log-uniform & Uniform & Uniform & Uniform & Uniform  \\ \hline \hline
    \end{tabular}
    \caption{Injected Parameters to Generate Synthetic Population and Priors used for Bayesian Inference}
    \label{tab:prior}
\end{table*}

\subsection{Eccentric Synthetic Population}
To make our study more realistic, we must add the measurement error in each source.  Rather than generate 
synthetic gravitational wave sources and perform full Bayesian inference, following previous work 
\cite{2017MNRAS.465.3254M} we generate mock measurement errors motivated by real parameter inference 
investigations. Hence, chirp mass and symmetric mass ratio are well-constrained compared to the primary and 
secondary masses of BBH. We compute them for each event in a population by using the following relation.

\begin{equation}
    M_c^T = \frac{(m_1 m_2)^{3/5}}{(m_1+m_2)^{1/5}},
\end{equation}

\begin{equation}
    \eta^T = \frac{(m_1 m_2)}{(m_1+m_2)^2},
\end{equation}
where $M_c^T$ and $\eta^T$ are found using the primary and secondary masses of each event generated by the 
power law model.
Furthermore, using the following relations, We add the measurement errors in the $M_c^T$ and $\eta^T$.

\begin{equation}
    M_c = M_c^T\left( 1+\beta (r_0+r ) \frac{12}{\rho}\right),
\end{equation}

\begin{equation}
\eta = \eta^T\left( 1+0.03 (r_0'+r') \frac{12}{\rho}\right),   
\end{equation}
where $r_0$ and $r_0'$ are the random numbers drawn from the standard normal distribution, which will shift 
the mean of the $M_c$ and $\eta$ distribution with respect to $M_c^T$ and $\eta^T$. The $r$ and $r'$ are the 
independent and identically distributed arrays of those randomly generated numbers to spread the distribution. 
The measurement uncertainty is inversely proportional to signal-to-noise ratio $\rho$, drawn from the 
distribution $p(\rho) \propto \rho^{-4}$, which holds for isotropically distributed sources in a static 
universe, subject to the threshold $\rho\geq 8$ for detection.   Following Section III.D of \cite{2022arXiv221007912W}, 
we estimate $\beta \simeq 0.5(v/0.2)^7/w$ where $v$ is an estimated post-Newtonian orbital velocity at a reference frequency of 20
Hz,  $w=\rho/12$, and  $\rho$ is drawn from a Euclidean SNR distribution $P(>\rho)\propto 1/\rho^3$.

Finally, after adding the measurement errors in the $M_c$ and $\eta$, we will convert them back to $m_1$ and 
$m_2$ to perform our analysis. We used the following relation for conversion, and it will provide the masses 
based on the condition $m_1\geq m_2$.
\begin{align}
    m_1 = \frac{1}{2} M_c \eta^{-3/5} (1+\sqrt{\eta_v}), \\
    m_2 = \frac{1}{2} M_c \eta^{-3/5} (1-\sqrt{\eta_v}), 
\end{align}
where $\eta_v = 1-4\eta$, we kept the samples with non-negative values and ignored the negative samples to 
avoid the square root issues. 

We also added the absolute error in the eccentricity distribution using the truncated normal distribution 
(to keep $\epsilon$ positive) scaling at $0.06$ and $0.2$ for modest and optimistic eccentricity case respectively.
The choices of distribution parameters are arbitrary for our study.

\subsection{Circular Synthetic Population}

To compare the synthetic EBBH population with the CBBH, we estimate how our sources would 
be characterized by parameter inferences that omitted the effects of eccentricity.  
Following \cite{favata-scaling-2022}, the effective chirp mass is well-constrained by observations dominated 
by the inspiral of a slightly eccentric binary:
\begin{align}
\label{eq:scaling}
M^{ecc} = \frac{M}{(1-\frac{157}{24}\epsilon^2)^{3/5}}
\end{align}
Our ansatz for source identification and characterization is that the best-fitting parameters and posteriors 
are directly related to the true posteriors, except that the recovered chirp mass is given by  
Eq. \ref{eq:scaling}.  This process removes the eccentric component from the population by scaling the masses 
and omitting the eccentricity. 

The significant effect after scaling is the mass shift, which can be observed in Fig. \ref{fig:pop2d_0.05_0.15}. 
In this figure, the left-hand side shows the mass shift of the first population generated with lower 
$\sigma_\epsilon =0.05$, which leads to a lesser mass shift. However, on the right hand of 
Fig. \ref{fig:pop2d_0.05_0.15}, one can see the significant mass shift after removing the eccentric component.

\section{Population Inference of EBBH and CBBH}
\label{sec:pop_inference}
\begin{figure*}[!htbp]
    \includegraphics[width=0.9\textwidth]{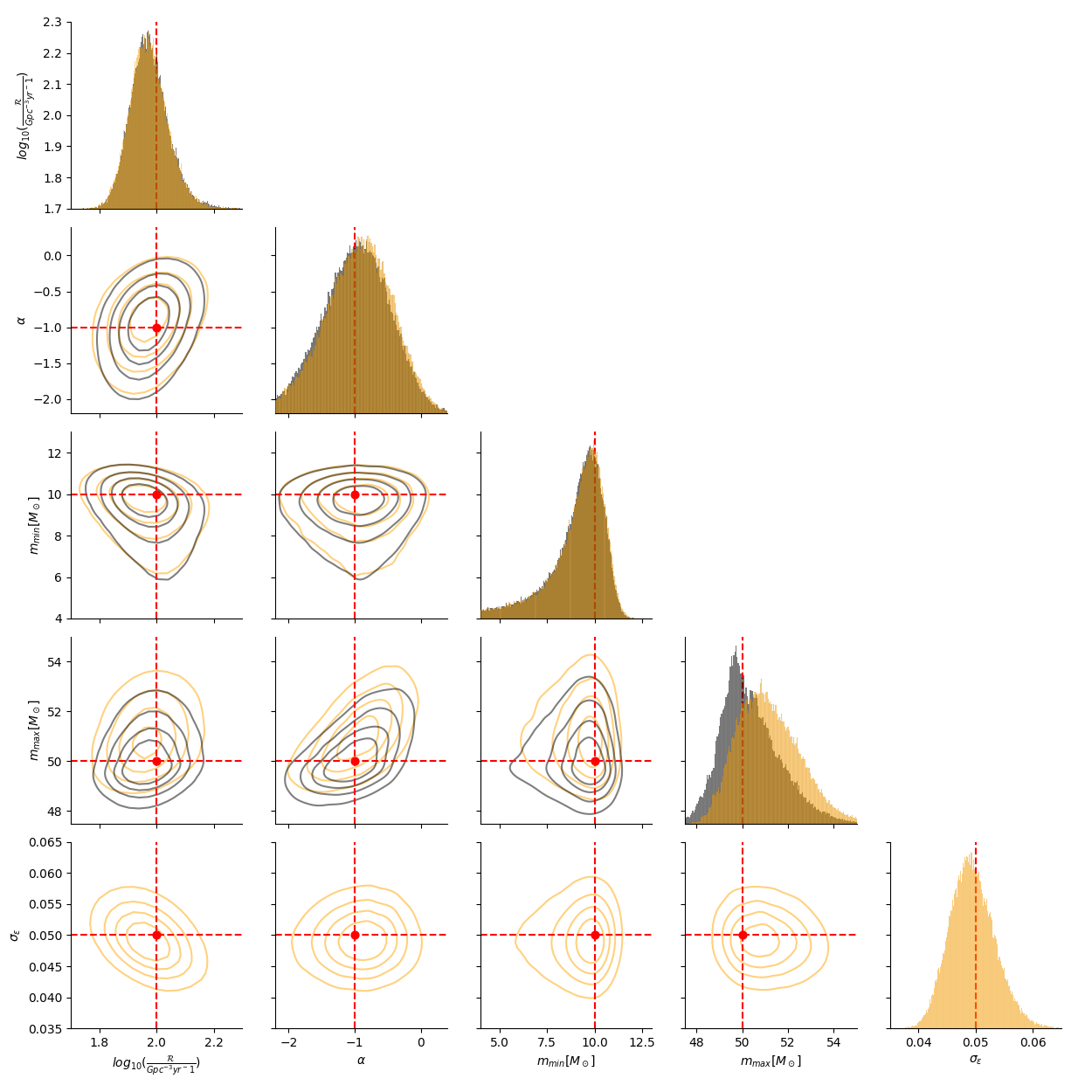}
    \caption{\label{fig:cor0.05}Corner Plots of EBBH (Orange) and CBBH (Black) for $\sigma_\epsilon = 0.05$}
\end{figure*}

\begin{figure*}[!htbp]
    \includegraphics[width=0.9\textwidth]{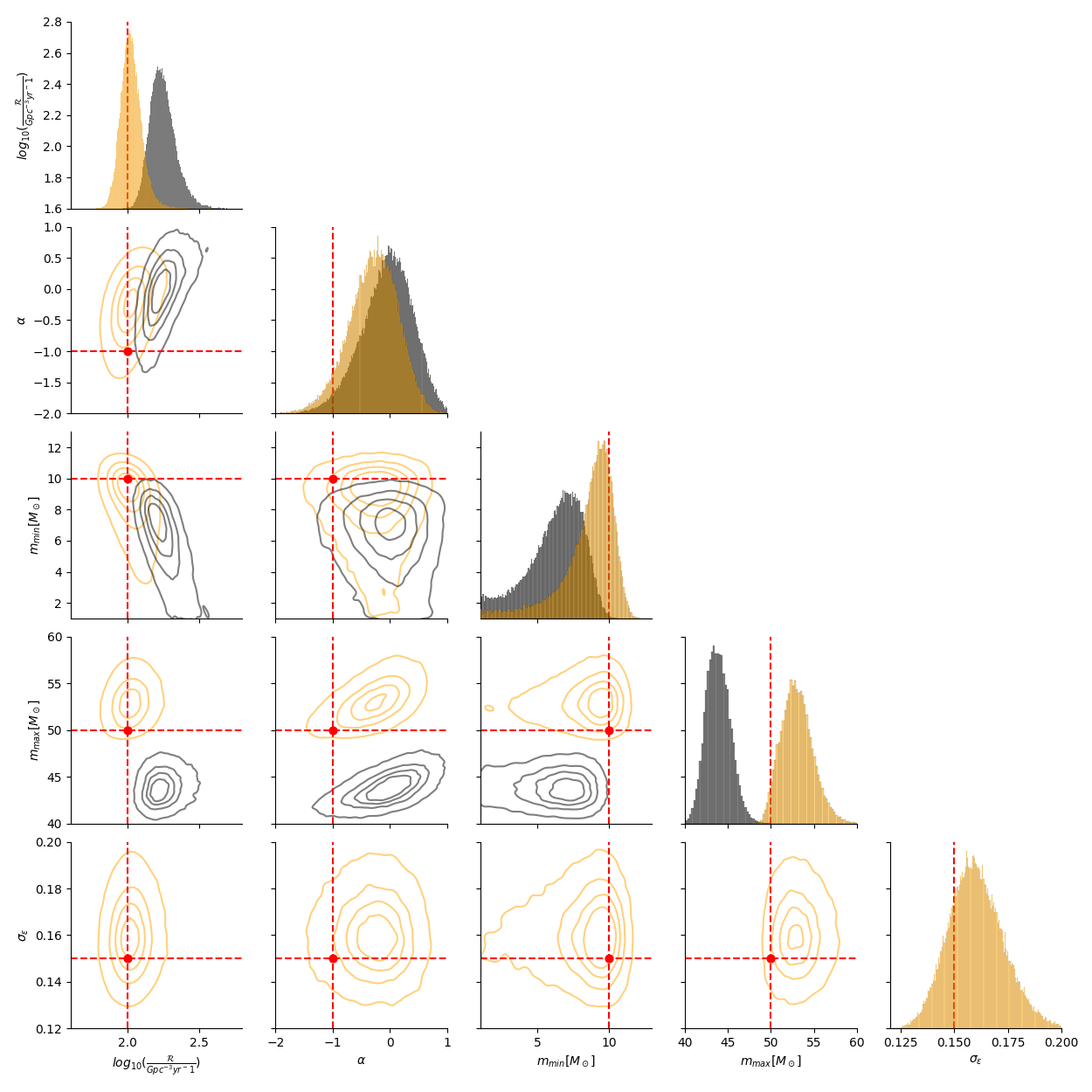}
    \caption{\label{fig:cor0.15}Corner Plots of EBBH (Orange) and CBBH (Black) for $\sigma_\epsilon = 0.15$}
\end{figure*}

We have generated two populations using conservative $\sigma_\epsilon = 0.05$ and optimistic 
$\sigma_\epsilon = 0.15$  to show the effect of eccentricity on the masses. We made the 
inference on both populations using Bayesian inference with uniform priors given in table 
\ref{tab:prior}, and we used a likelihood given in Eq. \ref{eq:likelihood_pop}. We estimated posterior 
distribution by collecting ample samples using  MCMC method. The contour plots of the posterior distribution 
for recovered parameters are given in Fig. \ref{fig:cor0.05} and \ref{fig:cor0.15} 
which shows the results of population inference under the most conservative  and optimistic case.
In those figure, the black color shows the results of CBBH, the orange color shows 
the results of EBBH, and the red color shows the true injected values.

The Fig \ref{fig:cor0.05} shows that in the limit of extremely small eccentricity, the impact of 
eccentricity is minimal on population constrains.  This 
result depends, of course, on our assumption that all binaries are equally likely to have eccentricity and 
that our measurement uncertainty in eccentricity is both independent of mass and fairly small, both optimistic 
assumptions. Most Notably, is the impact of neglecting eccentricity for higher eccentric binaries.  If the eccentricity is neglected, 
then the population inference is biased away from true values: the black contours in 
Fig. \ref{fig:cor0.15} do not include the true values for the injected population. This bias is
expected, as the eccentricity is a significant parameter in the population model particularly
at the higher eccentricities.
%


\begin{table*}
    \centering
    \begin{tabular}{c|ccccc}
        \hline \hline
        Inference & $log_{10}(\frac{\mathcal{R}}{Gpc^{-3}yr^-1})$ & $\alpha$ & $m_{min} [M_\odot] $ & $m_{max} [M_\odot]$ & $\sigma_\epsilon$ \\ \hline
        EBBH & $2.02^{+0.07}_{-0.07}$ & $-0.26^{+0.41}_{-0.45}$ & $9.03^{+1.13}_{-2.06}$ & $52.93^{+1.93}_{-1.78}$ & $0.16^{+0.01}_{-0.01}$ \\ \hline
        CBBH & $2.24^{+0.10}_{-0.08}$ & $-0.02^{+0.41}_{-0.47}$ & $6.56^{+1.54}_{-2.32}$ & $43.79^{+1.50}_{-1.34}$ & --\\ \hline \hline
    \end{tabular}
    \caption{Parameter Constrains: EBBH vs CBBH for $\sigma_\epsilon = 0.15$}
    \label{tab:inference_EBBH_vs_CBBH}
\end{table*}

\begin{figure}[!h]
    \includegraphics[width=0.45\textwidth]{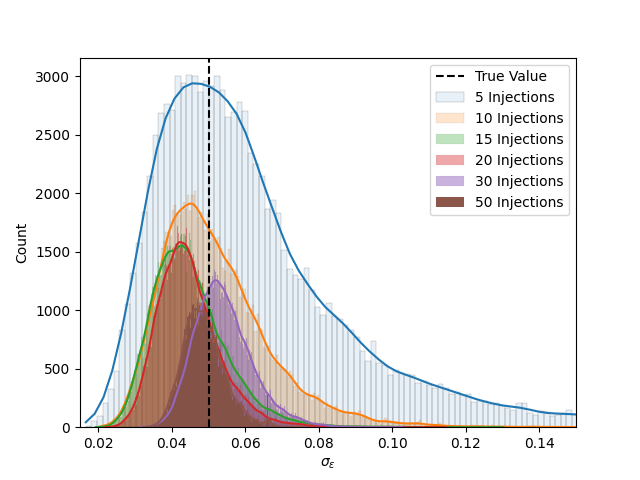}
    \caption{\label{fig:number} Histogram for eccentricity distribution with different numbers of injections.}
\end{figure}

In addition, we also performed our analysis to check the minimum number of eccentric events to recover the 
injected  $\sigma_\epsilon = 0.05$,  each with a measurement error of $0.05$. Our results show that we can 
recover the distribution of $\sigma_\epsilon$ even with five eccentric events. As we increase the number of 
events, our posterior distribution narrows toward the true value. So, the more eccentric events, the better 
the recovered value for the eccentricity distribution. 
The pattern of recovered eccentricity distribution from broader to narrow is evident in Fig. \ref{fig:number} 
with different numbers of events.

\section{Discussion}\label{sec:discussion}
The sensitivity of LVK detectors is increasing with time, leading to more detection in each observing run. 
We will have hundreds of events in future runs, and with the ongoing O4 run, detections could occur as 
frequently as every other day \cite{detection_rate_2016,detection_rate_2015}. 
Interestingly, the new detections lead us to the reasonable disagreement on the masses, spin, event rate, and 
formation scenarios \cite{LSC-BBH-2016, LSC-GW150914-2016}, which pushes researchers to develop new models 
\cite{Mandel-2016,marchant-2016}, to provide more information about the observable GW 
\cite{Barausse-2018, Abbot-2016-schotastic,dvorkin-2016}. Therefore, identifying and understanding the 
populations of EBBH with a growing number of detection will give us more insightful information about the 
formation and evolution of massive stars from birth to death over cosmic time.

Till now, all the confirmed detections are consistent with nearly circular orbits, and there can be various 
reasons for this bias. The first potential reason for getting circular orbits is that eccentric effects are 
more evident in low-mass events, and current searches are more efficient for higher-mass mergers. Secondly, 
it may result from selection biases in the waveforms because LVK detectors only use circular waveforms for 
parameter estimation (PE), which better present binaries evolved in an isolated environment than the 
dynamically evolved ones. 

Studies show that we must consider the multiple formation channels to understand the population better. Because 
a single channel may not contribute more than $70\%$ of the observation sample of BBH \cite{zevin-2021}. In 
addition, We may have stellar mass higher eccentric mergers at the lower frequency searches 
\cite{sesana-2016,chen-2017}, which can be measurable by detectors like Laser Interferometer Space Antenna 
(LISA) \cite{LISA-2017}. These observations would allow for long-term tracking of BBH orbital properties, 
which can be used to infer the formation mechanism better \cite{Breivik-2016}. So, it is critical for 
astrophysical implications to assess eccentricity distribution to infer their formation better.

\section{Conclusion}
\label{sec:conclude}
In this work, we demonstrate how one can recover the effects of eccentricity in a binary black hole population using a
parametric model.  Specifically, we extended a simple  power law model to include a one-sided Gaussian eccentricity
distribution for BBH. 
We verified our approach can recover the properties of two specific injected populations,  generating a large synthetic population 
with known values $log (\mathcal{R}) = 2$, $\alpha=-1$, $m_{min} = 10 M_\odot$, $m_{max} = 50 M_\odot$, and 
$\sigma_\epsilon = {0.05, 0.15}$. 
We then demonstrated that analyses using many fewer events could potentially identify the signatures of eccentricity. 
We expect $~1\%$ to $~17\%$ mass shift for modest $(\sigma_\epsilon = 0.05)$ 
to optimistic $(\sigma_\epsilon = 0.15)$ case respectively based on the Eq. \ref{eq:scaling}.
We observed $12.5\%$ difference in our population parameters recovery between EBBH and CBBH.
Finally, we demonstrated that the neglect of eccentricity can bias the recovered population parameters, if the
eccentricity in the population is frequently large evident in table \ref{tab:inference_EBBH_vs_CBBH}.




For this analysis, we restricted our inference only to BBH, but in the future, we aim to include eccentricity 
in other binary models, such as BNS and NSBH models. We also plan to modify the high-dimensional population 
models \cite{Dan_2019} to add eccentricity, spin, and precessing BBH.

\section*{Acknowledgements}

The authors acknowledge the computational resources provided by the LIGO Laboratory’s CIT 
cluster, which is supported by National Science Foundation Grants PHY-0757058 and PHY0823459.

\bibstyle{unsrt}
\bibliography{references.bib}

\end{document}